\documentclass[nocompress]{spie}  

\usepackage{amsmath,amsfonts,amssymb}
\usepackage{graphicx}
\usepackage[colorlinks=true, allcolors=blue]{hyperref}

\title{Theory of Ultrafast Spin--Charge Quantum Dynamics in Strongly Correlated
Systems Controlled by Femtosecond Photoexcitation: an Application to 
Insulating Antiferromagnetic Manganites}

\author[1]{P. C. Lingos} 

\author[2]{M. D. Kapetanakis} 

\author[2]{M. Mootz} 

\author[3]{J. Wang}

\author[2,1]{I. E. Perakis}

\affil[1]{ Department of Physics, University of Crete, Heraklion, Crete 71003, Greece and Institute of Electronic Structure and Laser, Foundation for Research and Technology-Hellas, Heraklion, Crete 71110, Greece}

\affil[2]{Department of Physics, University of Alabama at Birmingham, Birmingham, Alabama 35294, U.S.A}

\affil[3]{Ames Laboratory and
	Department of Physics and Astronomy, Iowa State University, Ames,
	Iowa 50011, U.S.A.}

\authorinfo{Further author information: Send correspondence to iperakis@uab.edu }

\begin{document} 
\maketitle

\begin{abstract}
We use a non-equilibrium many-body theory that engages the elements of transient coherence, 
correlation, and nonlinearity to describe changes in the magnetic and electronic phases of strongly
correlated systems induced by femtosecond nonlinear photoexcitation. 
Using a generalized tight--binding 
mean field approach based on Hubbard operators and including the coupling of the laser field, we describe a 
mechanism 
for simultaneous insulator--to--metal and anti-- to ferro--magnetic 
transition to a transient state triggered by non--thermal ultrafast spin and charge coupled excitations.
 We demontrate, in particular,  that photoexcitation of composite fermion quasiparticles induces quasi-instantaneous spin 
canting that quenches the energy gap of the antiferromagnetic insulator 
and acts as a nonadiabatic \lq\lq{initial condition}\rq\rq{} that triggers non-thermal lattice dynamics leading to an insulator to metal and antiferromagnetic (AFM) 
to ferromagnetic (FM) transitions. 
 Our theoretical predictions are consistent with recent ultrafast pump-probe 
spectroscopy experiments that revealed a magnetic phase transition   during 100fs
laser pulse photoexcitation of the CE--type AFM insulating phase of colossal magnetoresistive 
manganites. In particular, experiment observes 
two distinct charge relaxation components, fs and ps, 
with  
nonlinear threshold dependence  at a pump fluence
threshold that coincides with that for femtosecond magnetization photoexcitation.
Our theory attributes 
the correlation between femtosecond spin and charge nonlinearity leading to  transition in the magnetic and electronic state   to
spin/charge/lattice coupling and
laser-induced quantum spin canting 
that accompanies the driven 
population inversion between 
two quasi--particle bands with different 
properties: a mostly occupied polaronic band and a mostly empty metallic band, 
whose dispersion is determined by quantum spin canting. 
\end{abstract}

\keywords{Quantum Femtosecond Magnetism, Femtosecond Magneto-optical Pump-Probe spectroscopy, Non-equilibrium Photoinduced Phase Transitions, Colossal Magnetoresistive Manganites, Ultrafast Magnetization Dynamics, Optical Control of Magnetic Properties}

\section{INTRODUCTION}\label{sec:intro}

The spin-- and charge--ordered phases of quantum materials  can be switched nonthermally 
by using strong laser pulses to initiate non--equilibrium phase changes during fs timescales.
Non--equilibrium phase transitions  can be triggered  
 by laser--induced  charge fluctuations \cite{Chemla,Axt98} and by non--thermal  populations
of many--body states \cite{Karadim}, as well as by direct 
excitation of phonon oscillations. 
In this way, femtosecond (fs) laser pulses can   initiate non--adiabatic time evolution via different electronic, magnetic, and lattice pathways that proceed in parallel. "Sudden" fs photoexcitation  
 can create non-equilibrium transient states prior to the lattice thermalization that completes a phase transition \cite{Li-2013,SC,Lingos,Porer,Bigot}. 
Metastable phases arise at pre-thermalization timescales of a few picoseconds. 
Ultrafast pump--probe or THz spectroscopy experiments 
 can probe ultra-short times following strong pump photoexcitation
and access the initial temporal regime that drives such non-equilibrium laser--induced phase transitions. This  approach can be used to reveal
metastable hidden phases  prior to establishment of a new quasi--equilibrium
lattice structure.

Fully time-dependent theoretical and computational approaches are
required for addressing ultrafast spectroscopy experiments and for exploring non-thermal control schemes, including   
a quantum theory of 
coupled spin and charge 
dynamics in systems with deformable spin and lattice backgrounds. 
Such  approaches suitable for
addressing the non--equilibrium states observed in ultrafast spectroscopy experiments have lagged behind. Ginzburg-Landau equations, non-equilibrium dynamical
mean field theory, density matrix renormalization group, time-dependent density functional theory, exact
diagonalization, and canonical transformations all have advantages and disadvantages when modeling
time-dependent quantum many-body problems. For example, while time-dependent density functional
theory (TD-DFT) has successfully described molecular excitations, it faces difficulties when applied to extended
systems with large unit cells. Moreover, TD-DFT is too expensive 
computationally when addressing time-dependent quantum many-body
problems requiring hundreds of time steps and large super-cells containing hundreds to thousands of
atoms as in the manganites. 
 The
close proximity of many phases and the complex unit cells with many atoms, as well as disorder,
inhomogeneity, and lattice nonlinearities, limit the accuracy of ab initio methods when treating
the phase competition away from equilibrium, especially during non-thermal timescales. Further,
ab initio methods must include strong local correlations (e.g., LDA+U, LDA+DMFT) leading to
strong spatial variations.  Effective tight-binding
models \cite{Koskinen,Kotliar}
adequately describe energy bands close to the Fermi level, scale better for
calculations of large systems, and access longer timescales. Tight-binding density-functional
methods \cite{Koskinen} 
 can improve on the simpler models. 

In
semiconductors \cite{Chemla,Axt98,Karadim}, different theoretical methods have been developed that describe relaxation and
correlation by using density matrices derived from effective  Hamiltonians.
In the pre-thermalization temporal regime of interest here, photo-induced lattice displacements prior to the establishment of a new equilibrium lattice
structure change the electronic and magnetic structure. The latter is described to first approximation by using classical
equations of motion, with forces determined by the  electronic total energy,
including contributions from non-thermal electronic and spin populations, as well as from lattice
nonlinearities leading to metastability.  A modeling and computational approach
suitable for a large dynamic range in time
is needed in order to provide experimentalists with an
understanding of scaling and order parameters that lead to systematic understanding of the
ultrafast response of widely varying Phase Changing Materials types.
Such theory  aims to describe non-thermal
properties and metastable phases of quantum materials arising from many-body interactions
initiated by femtosecond laser pulses. 
Model  Hamiltonians 
studied with different theoretical techniques have been used 
to make progress in understanding  strongly correlated systems 
\cite{HubbardX}.
There is still a great deal of research
needed for theory to adequately describe observed experimental phenomena in phase changing materials.

 Femtosecond nonlinearities are known to lead to
non-adiabatic quantum dynamics in interacting electron-gas \cite{Karadim,Perakis}
and exciton \cite{Chemla,Axt98},
many-body systems.
While the well--studied femtosecond nonlinear response of
semiconductor nanostructures is determined by many-body interactions of carriers in “rigid”
energy bands, quasi-particle bands in quantum materials are “soft” and can thus be manipulated in different ways via  ultrafast laser pulses.
The
relevant Hamiltonians incorporate strong local interactions: Hubbard-U, spin-exchange, and
Jahn-Teller (JT) distortions. The
time-dependence arises from quasi-instantaneous
 populations of photo-generated carriers  and
nonlinear couplings of the strong laser fields. Non--equilibrium
methods for treating such quantum many body
Hamiltonians can be generally subdivided into
wave-function/density matrix and Green’s function
methods. Differences between these two
approaches arise from the handling of time propagation
in many-body systems. Non-equilibrium Green’s
functions require two time arguments, which strains
computer memory and is very time consuming.
Here we employ instead single-time density matrix methods
and treat correlations explicitly, separating the density matrix into quasi-thermal
and non-thermal parts \cite{Lingos}. The quasi-thermal component
  describes degrees of freedom that reach quasi--equilibrium
on timescales shorter than the experimental
resolution, while the non-thermal contribution is 
due to quantum kinetics of elementary excitations  prior to thermalization. 
The time--dependent part of the Hamiltonian drives non-adiabatic dynamics that generate non-thermal phase competition and
distinguishes fast from slow degrees of freedom. Ultrafast time dependence
arises from quasi-instantaneous coupling  of 
 quantum spin canting, charge fluctuations and correlations due to fs photo-excitation and quenching of electronic gap, as well by fast lattice displacements. 
To describe such effects, we  employ effective tight-binding models  
that can  describe bands close to the Fermi level, scale better for calculations of large systems,
and permit access to longer timescales.

The physical properties of complex materials such as colossal magnetoresisitive  manganites \cite{Dagotto} are governed by  collective order and fluctuations
of coupled degrees of freedom  \cite{SheuPRX,IchikawaNatMat,FiebigSci,PolliNat,RiniNat}.
This results  in elementary excitations and order parameters with coupled charge, orbital, spin, and lattice components, whose precise  microscopic composition remains unknown  \cite{Hubbard1,rucken,Milward,Aaron,Porer,Loktev}.
In the manganites, while  strong coupling of electronic, magnetic, and lattice 
degrees of freedom is believed to be responsible for the emergence  
of coexisting insulating/lattice--distorted/AFM and metallic/undistorted/FM nanoscale regions \cite{Dagotto}, 
the relevant quasi--particles have not been fully characterized yet \cite{Loktev}.  
Some theoretical studies have proposed that the sensitivity to the non--thermal electronic 
perturbations leading to the CMR phase transition from AFM/insulating to  FM/metallic state
may be due to delocalized electrons with mobility mediated by classical spin--canting \cite{Krish,Ramakrish,Milward}, which 
coexist with the polaronic carriers that dominate in the  insulating ground state \cite{Basov2011,kugel,millis}.  

In such quantum materials, there are two possible  pathways leading  to a photo-induced phase transition \cite{Wegkamp}: 
(1) an electronic charge and spin pathway triggered by laser-induced sudden excitation of non-thermal spin and charge populations. Subsequent relaxation of the photoinduced  electronic and spin populations 
also  changes the energy landscape  in the excited state, 
(2) a lattice pathway, which eventually leads to a delayed crystallographic phase transition that typically completes within ps timescales  \cite{subedi}. 
Here we are mostly interested in the first stage of the time--dependent process initiated by fs laser excitation of AFM 
insulating systems.
In  VO$_{2}$ \cite{Wegkamp} and TiSe$_{2}$ \cite{Porer} 
systems, previous experiments have shown that quasi-instantaneous electronic processes induced by the
 photo--carriers lead 
to metastable states prior to an insulator--to--metal phase transition, where the electronic and lattice orders evolve differently
\cite{Porer,Morrison}. 
For example, in the TiSe$_{2}$ insulator, the electronic component of
the charge density wave order parameter 
is quenched quasi-instantaneously while the lattice component persists
\cite{Porer}. 
This results in a non--equilibrium state  with lattice order similar to equilibrium,
whose properties are controlled by the photoinduced change in the local electronic density matrix 
 \cite{Porer}. 
In VO$_{2}$, a metastable metallic phase with the monoclinic lattice structure of the insulating phase is 
observed after the electronic component has switched from insulating to metallic  \cite{Morrison}.
So far, the role in such  non--equilibrium photoinduced states 
of spin  non--thermal fs dynamics 
is less understood. 
Here we discuss the role  of spin fluctuations driven by photo--carrier populations 
that interact with a deformable spin  and lattice medium. 

In low-bandwidth insulating manganites such as 
the Pr$_{0.7}$Ca$_{0.3}$MnO$_3$ system (PCMO),
metallic phases
cannot be accessed 
by tuning the temperature \cite{Dagotto}. 
An AFM insulator to FM metal equilibrium 
phase transition can, however, be induced  non--thermally,
e.g.  by applying a strong magnetic field, which leads to CMR.
While in equilibrium a magnetic field simultaneously changes the
coupled electronic, 
magnetic, and lattice order components, several ultrafast spectroscopy experiments
\cite{PolliNat,RiniNat,Beaud,Forst,Miyasaka,Matsubara,Okimoto,Wall,Ehrke,Matsuzaki,singla} 
 have  observed non--thermal  charge and/or spin dynamics prior to 
electron--lattice relaxation.
While  a new lattice structure 
seems to be established after ps timescales, electronic, orbital, and magnetic orders 
have been observed to change much faster.
Photo-induced  non-equilibrium phase transitions are typically
characterized   by a nonlinear threshold dependence
of the measured properties on the pump laser fluence. 
The time evolution of the 
charge, orbital, lattice, and magnetic  components of a complex order parameter 
can be separately monitored with fs X-ray diffraction (XRD) \cite{Beaud,Forst}. 
The fs  dynamics of AFM order is less
understood, 
as it may involve an AFM$\rightarrow$FM transition  initiated by charge excitations.  
Ref. \cite{Li-2013} reported 
a threshold increase of the  fs-resolved magneto-optical Kerr and
circular dichroism signals 
at 100fs time delays, which is absent at  ps time delays and only
appears below the AFM transition (Neel) 
temperature when a small magnetic field breaks the symmetry.
This  fs nonlinearity
 was interpreted 
in terms of an AFM$\rightarrow$FM  
transition that 
occurs  
prior to the ps spin--lattice relaxation \cite{Li-2013}.
It was  proposed 
that  
{\em quantum 
femtosecond magnetism} \cite{Bigot} 
and FM correlation may arise from 
both laser--driven 
charge fluctuations and non-thermal 
electronic populations.    
Here we discuss the 
 microscopic link between spin and charge nonthermal excitations in the fs temporal regime.

A possible nonthermal mechanism for photoinduced 
phase transition 
is based on the interplay between a quasi--instantaneous electronic/magnetic quantum pathway and 
a  lattice pathway. 
In particular, 
we discuss the possibility that the excitation of 
composite fermion itinerant quasi--particles dressed by spin fluctuations 
triggers quasi--instantaneously an insulator to metal and  AFM to FM 
phase transition. We 
 discuss  the  role  of  the 
 \rq{soft}  quasi--particle  energy bands, modified by fs laser excitation, which
  arise from  electron--magnon coupling 
 and the strong Hund's rule  local interaction. We note  that 
electron--magnon quantum fluctuations have been   observed experimentally 
to significantly change the spin--wave energy dispersion 
and lifetimes in  metallic 
manganites \cite{Kapet-corr,Kapet-SDW,SDW}.
For this purpose, we  present  a non-equilibrium modeling scheme
based on  quantum-kinetic equations of motion for the local electronic density
matrix coupled to lattice displacements. Our emphasis  here is on photo-induced quantum
many-body dynamics of intertwined charge, spin and lattice orders that govern
electronic, magnetic, and structural phase transitions in manganites. Such theory can  
facilitate discovery and identification of hidden
metastable quantum phases and emergent order, which is  accessed via laser-driven non-adiabatic pathways
and quantum fluctuations. 
We use a generalized tight-binding approximation  based on Hubbard operators \cite{Hubbard1},
with parameters can be  taken from the manganite literature \cite{Loktev,Dagotto},
to investigate how local correlations
affect quasi-particle spectra and collective spin behavior following ultrafast photo-excitation, by
extending the semiconductor Bloch equations \cite{Chemla} to strongly correlated systems. 
We model coupled, non-thermal spin-charge dynamics for itinerant electrons hopping between
clusters with localized electrons in well--defined spin states. We  treat these localized
electrons as core spins S with magnetic quantum numbers. 
Total
spin is conserved during electronic motion and we separate the J=S+1/2 and J=S-1/2 Hubbard bands. Itinerant
electrons preferentially form J=S+1/2 total spin states due to the strong Hund's rule interaction.
Electron motion is then restricted and
deforms the core AFM spin background via quantum fluctuations induced by spin-flip magnetic
exchange interactions.
We present numerical results for a  CE--type AFM 
unit cell. Our calculations show  
 strong coupling   
of the AFM chains and planes 
that characterize the insulating states of the manganites. 
In particular,  quantum spin fluctuations 
induced by the charge excitations
\cite{Dagotto,Ramakrish,Loktev}  
delocalize the  excited quasi--particles due to  deformation of the AFM
spin background during the electronic hopping. 
We show that this  leads  to a broad metallic  conduction band and a
quench of the insulator energy gap
facilitated by photoinduced FM correlation. 
As a result, the critical value of the Jahn--Teller (JT) lattice 
displacement necessary for stabilizing an insulating state 
changes. Using the above results, we propose a theoretical description of ultrafast  transition 
to a non--equilibrium metallic state with FM correlation 
and lattice displacements, which differs from  equilibrium.

\section{Problem Setup }\label{problemsetup}

An important characteristic of strongly correlated systems is the strong on--site 
local interactions, such as Coulomb, electron--phonon/Jahn--Teller, spin--exchange, etc-- 
whose strength well--exceeds the kinetic energy bandwidth. 
To first approximation, strong local interactions restrict the population of the local electronic configurations 
thatare energetically unfavorable, which restricts the itinerant electron motion. 
Rather than describing the dynamics in terms of bare electron operators, 
as in weakly--coupled semiconductors and metals, we start from the atomic limit 
and the Hubbard picture\cite{Hubbard2}. In this picture, 
 electronic motion arises from transitions between local many-electron configurations with given total spin.
We thus consider a lattice of 
atomic clusters located at sites labeled by $i$. We use the basis of the
atomic many--body states that diagonalize the local Hamiltonian at each lattice site. 
We assume that 
site $i$ can be populated 
by either $N$ or $N$+1 electrons, with the fluctuation in the population of the local configurations 
caused by an itinerant electron hopping from site to site or by laser photoexcitation. 
We assume that the atomic Hamiltonian of an \lq\lq{empty}\rq\rq{} site $i$ ($N$ localized electrons) has eigenstates $|im\rangle$, where $m=-S,\cdots S$ labels the z--component of a
local spin $S$ ($S_z$).
The limit $S=0$ corresponds to an empty site without local spin,
while the limit $S\to\infty$ corresponds to a classically treated local spin. 
The local  eigenstates at \lq\lq{full}\rq\rq{} sites populated by
a single itinerant electron ($N$+1 total electrons) are labeled by $|i\alpha M\rangle$, where M is the z--component of the total spin $J_i=S_i +s_i$ formed by
the itinerant and the $N$ localized electron spins.  Index $\alpha$ describes all other quantum numbers, such as different orbitals in the
presence of local (Jahn--Teller) lattice distortions, different total angular momentum values, doubly--occupied states
($N$+2 electrons), etc. 
We impose the local constraint
\begin{equation}
\sum_m |im\rangle \langle i m| +\sum_{\alpha M} |i\alpha M\rangle \langle i \alpha M| =1 
\label{complete} 
\end{equation}
at each individual site, which implies that 
only the population of the above local many--electron configurations can  significantly change  during the itinerant electron motion. 
  In the manganites, for instance, 
the relevant \lq\lq{active}\rq\rq{} Mn orbitals are the three--fold degenerate
$t_{2g}$--orbitals and the partially--filled (or empty) doubly--degenerate $e_g$--orbitals 
while the neighboring oxygen atoms participate as well similarly to the Zhang--Rice local singlet between the O hole and Cu$^{2+}$ ion in the Cu--oxide superconductors.\cite{Zhang}
The most important configurations correspond to total spin $J$=2 in "bridge" sites ($|i\alpha M\rangle$) and 
$S \approx$3/2 in \lq\lq{corner}\rq\rq{} sites ($|im\rangle$). The population of the two sites differs by $\sim 1$ in equilibrium, 
with the extra charge populating oxygen holes and $e_g$ manganite orbitals. 
Moreover,  the population of states with total spin $J=S-1/2$ is suppressed for  strong Hund's rule ferromagnetic interaction $J_H\to\infty$ 
between the itinerant and local electron spins on a given site.
Finally, the Jahn--Teller (JT) local
deformation of MnO$_6$ octahedra lifts the local state degeneracy  of the electronic states \cite{Dagotto,DagottoPRL,Tokura}.

\begin{figure} [t]!
   \begin{center}
   \begin{tabular}{c} 
   \includegraphics[scale=0.5]{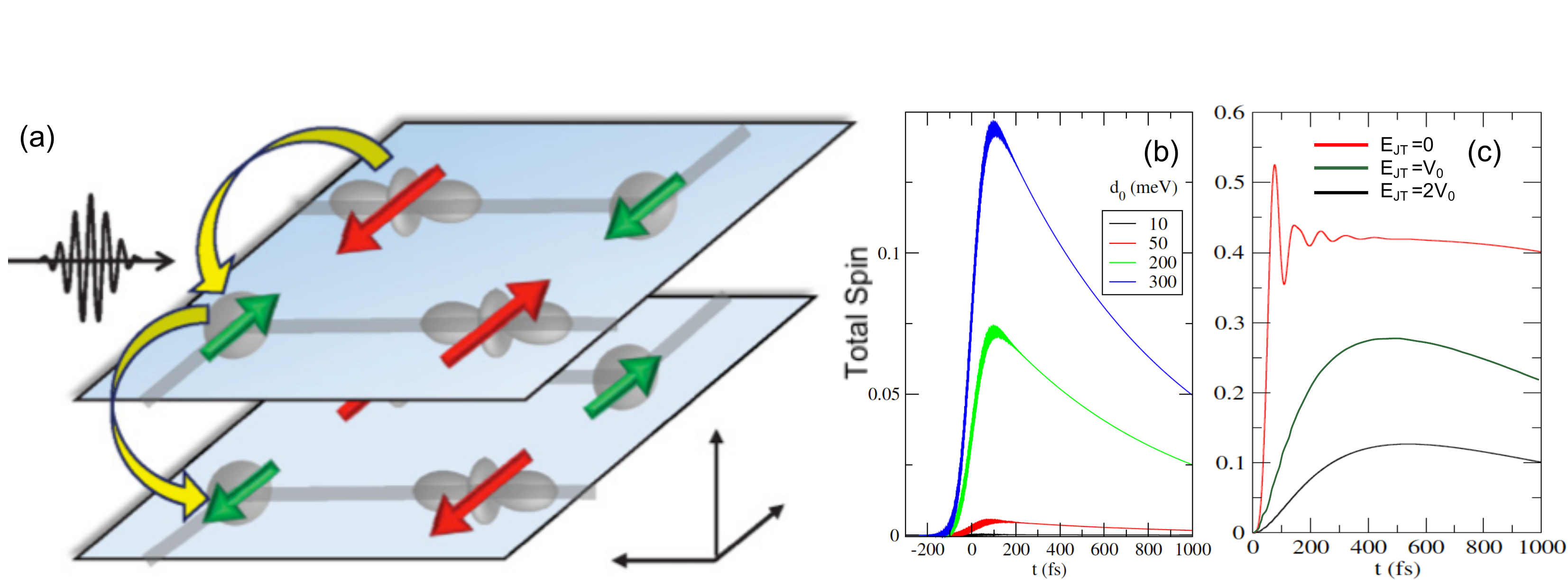}
   \end{tabular}
   \end{center}
   \caption[example] 
   { \label{fig1} 
(a) Illustration of the intersite electronic excitations in
the CE-AFM/CO/OO three-dimensional unit cell considered here,
which consists of 16 sites in two stacked AFM planes and two AFM--coupled
FM zigzag chains with alternating JT--distorted (bridge)
and undistorted (corner) sites. 
(b--c) 
 Calculated time--dependence of the total z--component
$S_z(t)$ of the two AFM core spins discussed in the text, driven by the
coupling of a 100-fs optical field pulse, for population lifetime $T_1=1$
ps and different Rabi energies $d_0$ (a), and (c) by the pulsed change discussed in the text, $V_0=100$ meV.
}
   \end{figure}

Here we are mostly interested on the role of 
spin fluctuations
in laser--induced insulator$\rightarrow$metal and AFM$\rightarrow$FM 
non--equilibrium phase transitions. 
Within classical spin scenarios, the  itinerant 
electrons  move on top of an adiabatically--decoupled 
spin background 
with their spins FM--locked 
to the  localized electron spins at each site: 
$M$=$S$+1/2 
and  $m$=$S$. 
Such an adiabatic picture assumes that the 
electronic hopping fluctuations occur on a time scale  faster than the spin dynamics. 
The  local spins then point along 
the   spin--canting angles $\theta_i$, which define a local z--axis 
that varies from site to site and defines the quasi--equilibrium 
directions of the local spins that change adiuabatically, 
i.e. much slower that quantum spin and charge fluctuations. 
For an equilibrium  state consisting of 
 AFM--coupled chains with FM spin correlation, as in the CE--AFM state of Fig. \ref{fig1}(a), 
spin conservation then restricts 
the electronic motion for large $J_H$, 
due to the magnetic exchange  energy cost for creating  an anti--parallel 
spin configuration 
\cite{Dagotto}. 
While such effects also determine the equilibrium properties, here 
we are interested in their pump--induced changes following 
strong pump photoexcitation of itinerant carriers. 
On the other hand,  quantum spin fluctuations  allow the non--equilibrium photoelectrons to  hop on sites with anti--parallel  spins
by flipping the localized  spins  via 
$J_H S^{\pm }_{i}\cdot s^{\mp }_{i}$.
They can then form 
states 
with
$J$
=$S$+1/2 but $M$=$S$-1/2 or smaller 
via  
electron--magnon quantum fluctuations  \cite{Kapet-corr,Kapet-SDW,SDW}. 

In the strongly correlated limit, we describe the electronic and spin excitations in terms of transitions between the
above atomic many--body states, created by local \lq\lq Hubbard operators\rq\rq. 
An example is the familiar Frenkel exciton
operator\cite{frenkel}, 
which creates a transition between the ground state and e--h excited state of an atom. 
Excitations that
conserve the total number of electrons are described in terms of the Hubbard operators
\begin{eqnarray} 
&& 
\hat{X}_i(m;m\prime)=| i  m \rangle \langle i  m^{\prime} |, 
\hat{X}_i(\alpha M;\alpha^{\prime} M^{\prime})=
| i \alpha M \rangle \langle i \alpha^{\prime}  M^{\prime} |, 
\end{eqnarray} 
which commute at two different lattice sites and satisfy bosonic--like non--canonical commutation relations. 
Our theory for the spin dynamics is based on
quantum kinetic equations of motion for the density matrix $\rho_i$ which is defined in terms of the populations of the local many--body
states at each site i as follows: 
\begin{eqnarray} 
&& 
\rho_i(m)=
\langle | i  m \rangle \langle i  m | \rangle, 
\rho_i^{\alpha}(M)=
\langle | i \alpha M \rangle \langle i \alpha  M | \rangle.
\label{rho-J} 
\end{eqnarray} 
Here we neglect photoinduced coherent couplings between the different local configurations 
$\alpha$. 
The above--defined local density matrix is  used here to describe the local spin at site $i$
and its quantum dynamics.  
Neglecting the $J$=$S$-1/2 high energy configurations, the  
core spin component
$S_z(i)$
along the local  z--axis defined by the 
equilibrium spin canting angle $\theta_i$ 
may be expressed 
as follows: 
\begin{equation} 
S_z(i)=
\sum_{m=-S}^S m \, \rho_i(m) + 
\sum_{M=-S-\frac{1}{2}}^{S + \frac{1}{2}}    
M \, \frac{ S}{S + \frac{1}{2}}  
\, \sum_{\alpha} 
 \rho^{\alpha}_i(M),
\label{spin-loc}  
\end{equation}  
Similarly, the z--component of the itinerant electron spin is expressed as 
\begin{equation} 
s^\alpha_z(i)= \frac{1}{2S+1} 
\sum_{M=-S-\frac{1}{2}}^{S + \frac{1}{2}}    
M \, 
 \rho^{\alpha}_i(M).
\label{spin-itin}  
\end{equation} 
 The z--axis of spin quantization refers to the equilibrium direction to
which the spin relaxes on a given site. It therefore varies from site to site. The above expressions for the spin on
site i were obtained after noting that the eigenstates of the magnetic exchange interaction $J_H \bf{S}_i\cdot\bf{s}_i$ coincide with the eigenstates of 
the total spin and, for $J=S$+1/2, are given by
\begin{equation}  \label{state1} 
|i \alpha M  \rangle =
\sqrt{\frac{S + M + \frac{1}{2}}{
{2S + 1}}} 
 \ | i \alpha; \uparrow  M-\frac{1}{2} \rangle + \sqrt{\frac{S - M + \frac{1}{2}}{2S+1}} \   \ |i \alpha; 
\downarrow 
M+\frac{1}{2} \rangle,
\end{equation}  
where $M$=$-J \cdots J$   and $\alpha$ 
labels the eigenstates  
of the JT   
and all other local interactions on site $i$. The above equation introduces the  Glebsch-Gordan coefficients 
\begin{equation}  \label{GG}
F_{\sigma}(M)=
\sqrt{\frac{S + \frac{1}{2} + \sigma M}{2S + 1}}.
\end{equation} 
Here by neglecting all $J$=$S-$1/2 
configurations, 
we do not include any spin dynamics on time scales of the order of the inverse magnetic exchange energy $\sim\hbar/J_H$, which
is typically in the sub--femtosecond range. 
We thus neglect the effects of the upper magnetic Hubbard band. 
Instead, we describe spin dynamics driven by the off-diagonal spinflip
interaction $J_H S_i^{\pm}\cdot s_i^{\pm}$,
which couples the different M states with the same magnetic energy (quantum spin flucutations). 
In equilibrium,  only the M=S+1/2 state is populated, similar to classical spins.  
However, as the
photoexcited electrons hop from site to site by 
flipping the local spins and exciting low energy magnons, different
spin  configurations can be populated off--equilibrium. 
For example, a spin--$\uparrow$ electron can hop on an empty site with local
spin pointing along the -z direction ($|i-S\rangle$ state) and form a $|i,\alpha,-S+1/2\rangle$ local configuration, Eq.(\ref{state1}), 
at zero magnetic energy cost. Such electron--magnon states are not allowed in the classical spin limit
$S\to\infty$. Their effect on the femtosecond spin dynamics is treated non--perturbatively here.

The equations--of--motion for the  local density matrix Eq.(\ref{rho-J}) 
at a given site $i$ 
couple to neigboring  inter--site coherences. The latter  
describe transient couplings  of the atomic quantum states in two different lattice sites, which can be driven 
by the laser field or by ultrafast changes in the hopping amplitudes.  
 To describe  inter--site electron hopping in 
a deformable spin background in 
the presence of strong local  correlations, we introduce \rq{composite fermion} local
electronic excitations created by Hubbard operators \cite{Hubbard1,Hubbard2}  that change the number of electrons 
 by one
and describe transitions between 
the active multi--electron configurations 
on a given site $i$: 
  \begin{equation} 
\hat{e}^\dag_{\alpha \sigma}(i M)
= | i \alpha M \rangle \langle i, M - \frac{\sigma}{2}|,
\label{Hubb-def1}
\end{equation}
where $\sigma$=$\pm$ 1 labels the z--component of the local excitation 
total spin ($\hbar$ =1). The above composite fermion operators create a total spin--$\sigma$/2
local excitation. 
The Hubbard operators obey the noncanonical 
 anticommutation relations \cite{Hubbard1,rucken}
\begin{eqnarray} 
[
\hat{e}^\dag_{\alpha^\prime \sigma^\prime}(i^\prime M^\prime),
\hat{e}_{\alpha \sigma}(i M) 
]_{+} 
= 
\delta_{i i^\prime} \
\left[
\delta_{M^\prime,M+\frac{\sigma^\prime-\sigma}{2}} 
\ | i \alpha^\prime M^\prime \rangle \langle i \alpha M|  
+ 
\delta_{M^\prime,M} \ \delta_{\alpha,\alpha^\prime} \  
| i, M-\frac{\sigma}{2} \rangle \langle 
i, M^\prime-\frac{\sigma^\prime}{2} | \right].
\label{fermi2}  
\end{eqnarray} 
The difference of the composite fermion anticommutators from
the usual fermion anticommutators is sometimes refered to as the \lq\lq{kinematic interaction}\rq\rq{} and arises from the strong
local correlations that restrict the populations of the different atomic many--body states. 
We thus project the
bare electron operators on the subspace of populated atomic states, Eq.(\ref{state1}), by using Eqs (\ref{spin-loc})--(\ref{spin-itin}) and the Glebsch--Gordan
coefficients Eq.(\ref{GG}) to obtain the 
projected Hamiltonian that conserves spin: 
\begin{eqnarray} 
 H(t)=\sum_i
\sum_{\alpha M} 
E_{i}(\alpha M) \,  |i \alpha M \rangle 
\langle i \alpha M| 
+ \sum_{i} \sum_{m}
 E_i(m) 
\,  | i m \rangle \langle i m|
 + H_{hop},
 \label{H(t)} 
\end{eqnarray}
The local excitation  energies 
\begin{equation} 
\varepsilon_{\alpha \sigma}(i) 
=
E_{i}(\alpha M) 
- E_{i}(M - \frac{\sigma}{2}),
\label{e-energ} 
\end{equation}
depend on the lattice coordinates
due to the electron--lattice (JT) local 
coupling, which lowers excitation energy, 
$\varepsilon_{\alpha \sigma}(i)=-E_{JT}(Q_i)$, where  $Q_i$ describes the local classical lattice distortion,   
while 
in all JT--undistorted sites, 
$\varepsilon_{\alpha \sigma}(i)$=0
if we neglect high energy electronic configurations $\alpha$. 

The quasi--particle inter--site hopping 
is described by \cite{Hubbard1} 
\begin{eqnarray} 
&& H_{hop}(t)=-  \sum_{i i^\prime}  
\sum_{\sigma}  
\sum_{\alpha \alpha^\prime} 
V_{\alpha \alpha^\prime}(i- i^\prime) 
 \left[ \cos \left(\frac{\theta_i - \theta_{i^\prime}}{2} \right) 
\, \hat{e}^\dag_{\alpha \sigma}(i) \,
\hat{e}_{\alpha^\prime \sigma}(i^\prime)
+ \sigma \, 
 \sin \left(\frac{\theta_i - \theta_{i^\prime}}{2} \right) 
\, \hat{e}^\dag_{\alpha \sigma}(i) \,
\hat{e}_{\alpha^\prime -\sigma}(i^\prime) \right],
\label{hop}
\end{eqnarray}
where  
$\hat{e}^\dag_{\alpha \sigma}(i)= \sum_M F_{\sigma}(M) \hat{e}^\dag_{\alpha \sigma}(i M)$
 and 
the amplitudes
$V_{\alpha \alpha^\prime}(i- i^\prime)$
have both static ($t_{\alpha \alpha^\prime}$) 
and laser--induced ($\Delta V_{\alpha \alpha^\prime}$) contributions, 
$V_{\alpha \alpha^\prime}(j-i)= 
t_{\alpha \alpha^\prime} + 
 \Delta V_{\alpha \alpha^\prime}(t)$.
 Importantly,  Eq. (\ref{hop}) ensures  conservation of total spin 
during electronic motion. 
Transient changes in the above inter--site hopping 
amplitudes, $\Delta V$, can arise from either the direct coupling 
of the optical field
or by photoinduced transient changes in the local lattice distortions $Q$
and local many--electron configurations $|i \alpha M\rangle$. 
The above Hamiltonian is quite general and 
does not rely on the details of the local configurations at each site $i$. The latter 
 determine the effective parameters $E_{i}(\alpha M)$ and $V_{\alpha \alpha^\prime}(i - i^\prime)$, which can also be fitted 
 from a more micrtoscopic theory. 

In the classical spin limit,
the only populated states have 
$m=S$ or $M=S+1/2$, 
as all spins  point along directions 
$\theta_i(t)$ that depend on the lattice site.
Introducing the deviation of $S_z(i)$ 
from its classical value, $\Delta S_z(i) = S -S_z(i)$,  
and using the completeness relation  Eq.(\ref{complete}) 
we obtain from Eq.(\ref{spin-loc}) 
 \begin{eqnarray} 
  \frac{ \Delta S_z(i)}{S}  =   
 \sum_{\alpha}  \sum_{M=-S-\frac{1}{2}}^{S - \frac{1}{2}}    
 \,  \frac{ S + \frac{1}{2} - M}{S + \frac{1}{2}}  
\,
 \rho^\alpha_i(M) 
 + \sum_{m=-S}^{S-1} \frac{S - m}{S} 
\, \rho_i(m). 
\label{DS}  
\end{eqnarray} 
The above equation  describes  canting  from the classical spin direction
$\theta_i$ due to  the population of 
local states with 
$M \le S - 1/2$ and $m \le S-1$  described by the 
local density matrix
Eq.(\ref{rho-J}).  
Therefore, the femtosecond quantum kinetics of the local (onsite) density matrix driven by the coupling of the 
laser field pulse, described below, 
introduces local spin quantum dynamics
that can lead to transient non--thermal changes in the magnetic interactions. 
Similarly, the local density matrix is used to describe 
the 
quantum fluctuations of the itinerant electron spin,
\begin{equation}\label{Ds-itin}
\Delta s^\alpha_z(i)= \frac{f_i^\alpha}{2} -  s^\alpha_z(i)
\end{equation}
where  
\begin{equation} 
f_i^\alpha 
= \sum_{M} \rho^{\alpha}_{i}(M),
\label{fi}
\end{equation}  
is the electron charge population 
on site $i$, which describes the classical contribution.

We describe the non--thermal spin dynamics introduced by "sudden" laser excitations 
using quantum kinetic  equations of motion 
for the spin--dependent local populations, obtained from  Eq. (\ref{H(t)}): 
\begin{eqnarray} 
\partial_t  \rho^{\alpha}_i(M) &=&
2 \, Im \,
 \, 
\sum_{\sigma^\prime=\pm 1 }  
F_{\sigma^\prime}(M) \times 
\sum_{l
\alpha^\prime  }V_{\alpha^\prime \alpha}(l-i) \times\nonumber \\
&&\langle 
 \Bigg[ 
 \cos\left(\frac{\theta_{l} - \theta_i}{2} \right)
\hat{e}^{\dag}_{\alpha^\prime \sigma^\prime}(l) 
 - 
\sigma^\prime  
 \sin \left(\frac{\theta_{l} - \theta_i}{2}\right)
\, \hat{e}^{\dag}_{\alpha^\prime  -\sigma^\prime}(l ) 
\Bigg]
\hat{e}_{\alpha \sigma^\prime}(i M) \rangle
 \label{dm-Xab-1} 
\\
\partial_t  \rho_i(m) 
&=&-
2 Im \, \sum_{\alpha} \, \sum_{\sigma^\prime} 
F_{\sigma^\prime}(m+\frac{\sigma^\prime}{2}) 
\, 
\times 
 \sum_{l \alpha^\prime}  
V_{\alpha^\prime \alpha}(l-i)\times\nonumber \\
&&\langle  \Bigg[ 
\cos\left(\frac{\theta_{l} - \theta_i}{2} \right) 
\,  \hat{e}^{\dag}_{\alpha^\prime \sigma^\prime}(l) 
- \sigma^\prime \,
 \sin\left(\frac{\theta_{l} - \theta_i}{2} \right) 
\, 
\hat{e}^{\dag}_{\alpha^\prime -\sigma^\prime}(l) 
\Bigg]  \hat{e}_{\alpha \sigma^\prime}(i, m+\frac{\sigma^\prime}{2}) 
\rangle. 
\label{dm-XM-1} 
\end{eqnarray}
The above equations  
 describe the  coupling of  site  $i$ 
to the 
rest of the lattice, which is  
driven by $H_{hop}(t)$, Eq.(\ref{hop}).
This time--dependent coupling is 
characterized 
by intersite
coherent 
amplitudes $\langle \hat{e}^{\dag}_{\beta \bar{\sigma}}(j)
\, \hat{e}_{\alpha \sigma}(i M) \rangle$
describing spin--dependent 
charge fluctuations. The latter inter--site coherences   
 are obtained from their equations of motion
 after using the 
factorization 
\begin{eqnarray}
&& \langle 
  [ \hat{e}^\dag_{\beta \bar{\sigma}}(j),
\hat{e}_{\beta^\prime \sigma^\prime}(j)
]_+ \, \hat{e}^\dag_{\alpha^{\prime} \sigma^{\prime}}(l) \, 
  \hat{e}_{\alpha \sigma}(i M) \rangle=
\langle 
  [ \hat{e}^\dag_{\beta \bar{\sigma}}(j),
\hat{e}_{\beta^\prime \sigma^\prime}(j)
]_+ 
\rangle 
\langle 
 \hat{e}^\dag_{\alpha^{\prime} \sigma^{\prime}}(l) \, 
  \hat{e}_{\alpha \sigma}(i M) \rangle,
\label{factorize} 
\end{eqnarray} 
where $j \ne l,i$.
The above mean field approximation neglects any  fluctuations in the composite fermion 
anticommutator $[ \hat{e}^\dag_{\beta \sigma}(j),
\hat{e}_{\beta^\prime \sigma^\prime}(j)
]_+$ 
as in the Hubbard--I approximation 
\cite{Hubbard1,Hubbard2}.
We  thus obtain after some simple algebra the equation of motion describing the 
time--dependent coupling between site $i$ and the rest of the lattce 
 $j \ne i$ 
  \cite{Lingos_PRB_2017}
\begin{eqnarray} 
&& i \partial_t  \langle \hat{e}^{\dag}_{\beta \bar{\sigma}}(j)
\, \hat{e}_{\alpha \sigma}(i M) \rangle
-\left [\varepsilon_{\alpha \sigma}(i)  
-\varepsilon_{\beta \bar{\sigma}}(j) 
\right] \, \langle \hat{e}^{\dag}_{\beta  \bar{\sigma}}(j)
\, \hat{e}_{\alpha \sigma}(i M) \rangle  
\nonumber \\
&&
 =
F_{\sigma}(M) \ \langle [ \hat{e}^\dag_{\alpha \sigma}(i M),
\hat{e}_{\alpha \sigma}(i M)
]_+
\rangle \ \sum_{l }  
\sum_{\beta^{\prime}}
V_{
\alpha \beta^\prime}(i-l)  \, 
\sigma \,
\sin\left(\frac{\theta_{l} - \theta_{i}}{2} \right) 
\, 
\langle  \hat{e}^{\dag}_{\beta \bar{\sigma}}(j) \,  
\hat{e}_{ \beta^{\prime}- \sigma}(l) \rangle 
\, 
\nonumber \\
&&
 -F_{\sigma}(M) 
\, \langle  [ \hat{e}^\dag_{\alpha\sigma}(i M),
\hat{e}_{\alpha \sigma}(i M)
]_+ \rangle\, 
 \sum_{l}  
\sum_{ \beta^{\prime}}
V_{
\alpha \beta^\prime}(i-l)  \,
\cos\left(\frac{\theta_{l} - \theta_{i}}{2} \right) 
\, \langle  \hat{e}^{\dag}_{\beta \bar{\sigma}}(j) \,  \hat{e}_{ \beta^{\prime} \sigma}(l) \rangle 
 \
\nonumber \\
&&
+\langle 
  [ \hat{e}^\dag_{\beta \bar{\sigma}}(j),
\hat{e}_{\beta \bar{\sigma}}(j)
]_+ \rangle \,
\sum_{l} 
\sum_{\alpha^\prime}
V_{
\alpha^\prime \beta}
(l-j) \, 
\cos\left(\frac{\theta_{l} - \theta_{j}}{2} \right) 
\, 
 \langle \hat{e}^\dag_{\alpha^{\prime} \bar{\sigma}}(l) \, 
  \hat{e}_{\alpha \sigma}(i M) \rangle 
\nonumber \\
&&
-\langle 
  [ 
\hat{e}^\dag_{\beta \bar{\sigma}}(j),
\hat{e}_{\beta \bar{\sigma}} (j)
]_+ \rangle \ \sum_{l} 
\sum_{\alpha^\prime}
V_{
\alpha^\prime \beta^\prime}
(l-j) \, \bar{\sigma}
\, \sin\left(\frac{\theta_{l} - \theta_{j}}{2} \right) 
\, 
\, 
\langle \hat{e}^\dag_{\alpha^{\prime} -\bar{\sigma}}(l) 
\, 
  \hat{e}_{\alpha \sigma}(i M) \rangle.
\label{eom-coh-Fourier} 
\end{eqnarray} 
Noting Eq.(\ref{fermi2}), 
the first two lines in the above equation 
describe the coupling 
to the local density matrix at site $i$. The last two lines describe coupling of  site $i$ 
to the rest of the lattice.   
The coherent long--range coupling 
between any two lattice sites $j \ne l$
due to itinerant electron motion  
is similarly described by the equation 
of motion 
 \begin{eqnarray} 
&& i \partial_t  \langle \hat{e}^{\dag}_{\beta \bar{\sigma}}(j)
\, \hat{e}_{\alpha \sigma}(l) \rangle  
-\left [\varepsilon_{\alpha \sigma}(l)  
-\varepsilon_{\beta \bar{\sigma}}(j)  
\right] \, \langle \hat{e}^{\dag}_{\beta \bar{\sigma}}(j)
\, \hat{e}_{\alpha \sigma}(l) \rangle  
\nonumber \\
&&
=  \langle  [ \hat{e}^\dag_{\beta \bar{\sigma}}(j),
\hat{e}_{\beta \bar{\sigma}}(j)
]_+ \rangle 
\sum_{l^\prime} 
\sum_{\alpha^\prime}
V_{
\alpha^\prime \beta}
(l^\prime-j) 
\ \langle 
 \left[ 
\cos\left(\frac{\theta_{l^\prime} - \theta_{j}}{2} \right) 
\hat{e}^\dag_{\alpha^{\prime} \bar{\sigma}}(l^\prime) 
- \bar{\sigma} 
 \sin\left(\frac{\theta_{l^\prime} - \theta_{j}}{2} \right)  
 \hat{e}^\dag_{\alpha^{\prime} -\bar{\sigma}}(l^\prime) \, 
  \right]  \hat{e}_{\alpha \sigma}(l) \rangle \
\nonumber \\
&&
 -  
  \langle  [ \hat{e}^\dag_{\alpha \sigma}(l),
\hat{e}_{\alpha \sigma}(l)
]_+ \rangle \ \sum_{l^\prime} 
\sum_{\beta^{\prime}}
V_{
\alpha \beta^\prime}(l-l^\prime) \ 
\
\langle 
\hat{e}^{\dag}_{\beta \bar{\sigma}}(j) 
\left[ \cos\left(\frac{\theta_{l^\prime} - \theta_{l}}{2} \right) 
 \hat{e}_{ \beta^{\prime} \sigma}(l^\prime) 
- \sigma 
\sin\left(\frac{\theta_{l^\prime} - \theta_{l}}{2} \right) 
\hat{e}_{ \beta^{\prime} - \sigma}(l^\prime) 
\right]  \rangle.
\label{HF-coh} 
\end{eqnarray}
The above equations 
of  motion  provide a closed 
system 
for  describing the laser--induced nonthermal time evolution 
of the local  
density matrix and spins.   
They also describe time--dependent chaNges in the effective  inter--site magnetic  interactions 
mediated by ultrafast charge fluctuations driven by the laser field 
or by photoinduced ultrafast "sudden" changes in the effective local excitation  
energies and inter--site hoppings (``bonding order"). 
In the next section we discuss how short--range FM correlation can be transiently  modified away from equilibrium 
by laser--induced charge fluctutations 
across the JT energy barrier..

The main difference between bare electrons and the composite fermion quasi--particles 
considered here 
comes from the 
 spin--dependent 
anti--commutators
\begin{eqnarray} 
&& n_{\alpha \sigma}(i)=\langle [
\hat{e}^\dag_{\alpha \sigma}(i),
\hat{e}_{\alpha \sigma}(i) 
]_{+} \rangle.
\label{electr-anticomm}
\end{eqnarray}  
The latter 
deviate from their fermionic values
due to the strong local correlations, here  
the suppression of 
the population of 
$J$=$S-1/2$ total spin configurations  
during the electronic motion.
Using the completeness relation  Eq.(\ref{complete}) 
and  Eqs.(\ref{spin-loc}) and (\ref{spin-itin}), we obtain 
\begin{eqnarray} 
n_{\alpha \sigma}(i) =
\frac{1}{2 S + 1}  
\left[ S + \frac{1}{2} + 
\sigma \left( S_z(i) + s_z^\alpha(i)
+ \frac{\sigma}{2} \left( 1 - f_i^\alpha \right) 
\right) \right].
\label{factor1}
\end{eqnarray} 
Introducing the quantum spin fluctuations, $\Delta J_z^\alpha(i) 
=\Delta S_z(i) + \Delta s_z^\alpha(i)$, 
Eqs. (\ref{DS}) and (\ref{Ds-itin}),
we obtain from the above equation 
\begin{eqnarray}  
n_{\alpha \sigma}(i) =
\frac{S + \frac{1}{2} + 
\sigma \left(S + \frac{1}{2} - \Delta J_z^\alpha
\right)}{2S+1} 
+
\frac{1 -\sigma}{2} \
\frac{1 - f_i^\alpha}{2S+1}.
\label{factor2}
\end{eqnarray}
In the limit of 
classical spins,  $S \rightarrow \infty$, 
Eq.(\ref{factor2}) gives 
$n_{\alpha \uparrow}(i)=1$ and 
$n_{\alpha \downarrow}(i)=0$ 
as in the case of bare electrons. 
In this approximation,  the electrons are effectively 
spinless, 
as their spin is locked with the 
core spins during their  motion, in a FM configuration 
parallel to  $\theta_i$ due to Hund's rule 
\cite{anderson,Dagotto}.
On the other hand, 
in the case of composite fermions, 
Eq.(\ref{factor2}) 
gives 
\begin{equation}
n_{\alpha \uparrow}(i) 
= 1 - \frac{\Delta J_z^\alpha}{2 S + 1} 
\ , \
n_{\alpha \downarrow}(i) 
= \frac{1 - f_i^\alpha + \Delta J_z^\alpha}{2 S + 1}  \ne 0, 
\label{factor} 
\end{equation} 
which allows for $\sigma$=-1 quasi--particle excitations 
with total spin anti--parallel 
to the equilibrium spin direction $\theta_i$.
In the above equation, the composite fermion 
anticommutators depend on the 
local spin fluctuations 
$\Delta J_z^\alpha$, as well as 
the filling factors $f_i^\alpha$ 
that determine the spatial modulation of the local 
charge. 

\section{Laser--Driven Ferromagnetic Correlation: Femtosecond quantum  spin dynamics } \label{two-sites}

\begin{figure}
   \begin{center}
	\includegraphics[scale=0.9]{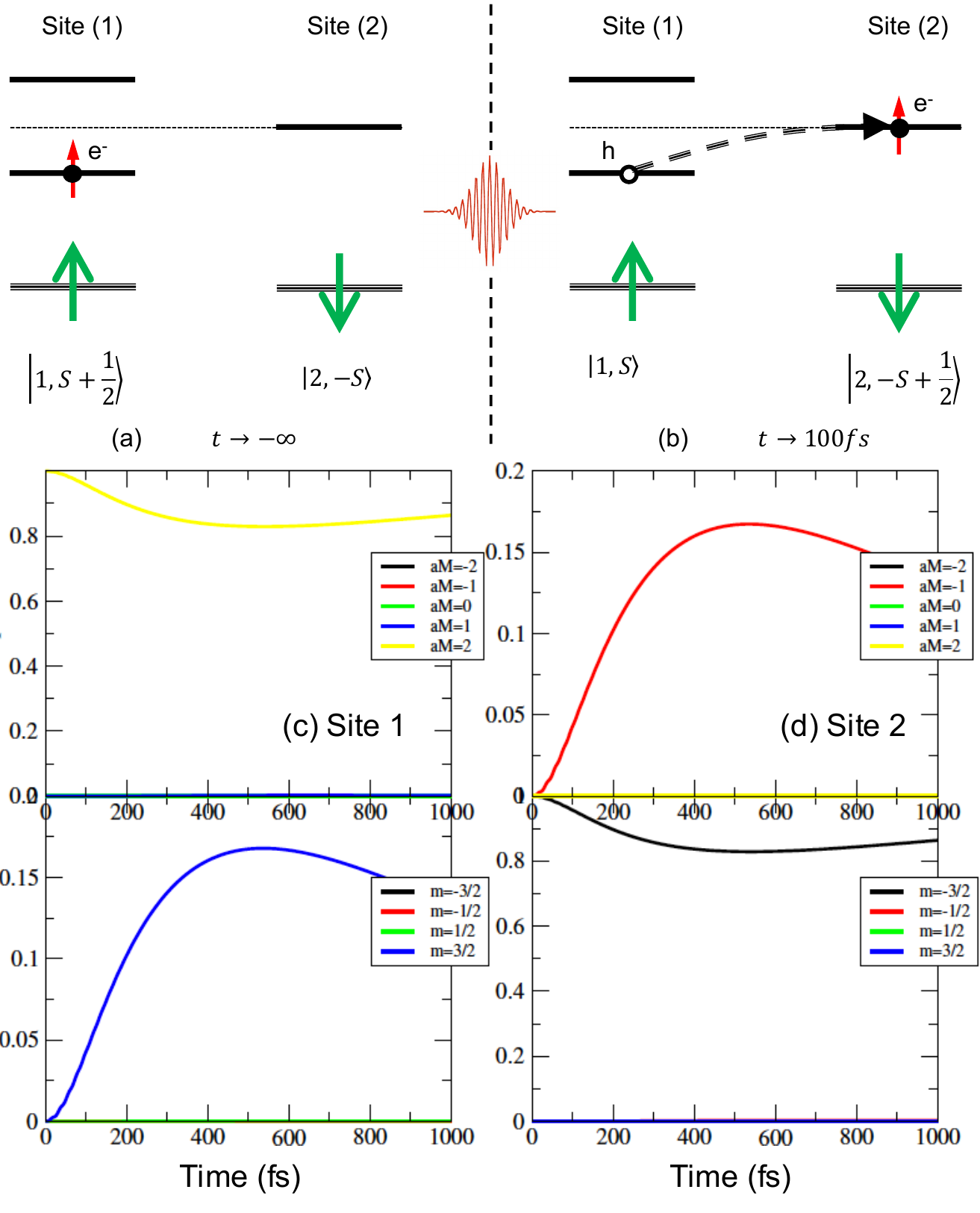}
   \end{center}
   \caption
   { \label{fig2} 
(a)--(b) Schematic of the dimer system before (a) and after (b) being driven by the optical field. 
(c)--(d) Time evolution of all photoinduced populations at the two sites  of the dimer system.
}
   \end{figure}

In this section we discuss an example of how laser--induced  charge fluctuations can quasi--instantaneously  excite
 spin dynamics and   short--range FM correlation between AFM sites based on quantum spin flucutations.    
To illustrate this possibility,
we consider 
a \lq\lq{quantum  dimer}\rq\rq{} of 
AFM local spins, which 
consists 
of a
JT--distorted site  (site 1) 
populated in the equilibrium state by a $J$=S+1/2
configuration $|i \alpha M\rangle$ 
with  
total spin  $M$=S+1/2  
and energy $-E_{JT}$, and 
an
undistorted site (site 2) 
populated by a  total 
core spin 
$S$ with z--component 
$m$=$-S$ anti--parallel to the spin  
at site 1 and energy zero (see Fig. \ref{fig2}(a)).  
We assume that the optical 
field induces   an ultrafast change in the intersite 
hopping amplitude $V$ from its equilibrium value.
Such time--dependent change in the hopping amplitude can be directly driven by the coupling of the 
optical field, which induces charge fluctuations between the two sites. 
However, it can also come from the excitation of non--thermal populations that 
change the local multi--electron configurations, by introducing, e.g., non--thermal lattice distortions
that last during 100fs non--thermal time scales and proportional to the non--thermal electronic populations
as discussed below \cite{zeiger}. 
The  approximation of fairly localized  charge density \cite{dimers} simplifies the calculation 
and already captures some of the 
properties of the  extended system 
\cite{Krish,veenendaal,Loktev}.
It  leads to 
an effective Hamiltonian with
short--ranged   interactions   \cite{Loktev,Ramakrish,Krish}. Here, such  interactions  are modified by any
time--dependent pulsed change in the hopping amplitude,  
 which  drives inter--site non--equilibrium charge fluctuations  as   
illustrated by the dash-lined arrow in Fig. \ref{fig2}(b).
Such charge fluctuations result in laser--induced non--equilibrium changes in the 
local density matrix, which also affect the
itinerant  quasi--particle 
dispersion, energy bandgap, and phonon properties  in correlated systems with 
\lq\lq{soft}\rq\rq{} energy bands. Previous examples include TiSe$_{2}$ \cite{Porer} or VO$_{2}$ \cite{He}.
 We  solve the quantum--kinetic 
equations of motion 
for the spin--dependent  density matrix 
of the two sites
 and then calculate
the 
 z--component of the total core spin
of the two above sites, $S_z$=$S_z$(1)+$S_z$(2), by using Eq.(\ref{spin-loc}).
We consider the same z--axis for both sites.
The local populations are assumed to relax on a timescale  
$T_1$=1ps, with the inter--site charge fluctuations  dephasing on a timescale 
$T_2$= 50fs.
However, the precise values of $T_1$ and $T_2$ do not change the qualitative behavior. 
Fig.1 compares the results for two different cases: (1) Optical field 
$\Delta V = V_0 e^{- (t/\tau)^2} e^{-i \omega t}$,
(2) Pulsed change  $\Delta V = V_0 (1 - e^{-t/\tau_1}) e^{-t/\tau_2}$, with $\tau_1$=100fs and 
$\tau_2$=500fs. 
In the latter case, we compare the time--dependence for $E_{JT}$=0, $V_0$, and =2 $V_0$. 
 $E_{JT}$=0 corresponds to hopping between two sites
with the same lattice distortion,  as is the case for hopping between two neighboring AFM planes
with the same lattice and charge configurations,
while $E_{JT}>0$ corresponds to hopping between a JT--distorted and undistorted  sites. 
Fig. \ref{fig1}(b) shows that a  finite $\Delta S_z(t)$ 
develops with time, which corresponds to 
a photoinduced FM correlation induced by the quantum spin fluctuations. 
Note that, for classical spins, charge fluctuations between the two AFM sites are prohibited for $J_{H} \rightarrow \infty$
considered here, so any spin dynamics is adiabatic, determined by quasi--thermal changes in the local canting angles 
$\theta_i$ that describe the spin background adiabatically decoupled from the electronic motion. 
A long--range  magnetization could  then
 arise  if a macroscopic 
number of such dimers (or clusters)  orients along a preferred direction
\cite{Li-2013,Matsubara,Ferrari}.
As seen in Fig. (\ref{fig1}), in all cases the total spin 
$\Delta S_z(t) = \Delta S_z(1) 
+ \Delta S_z(2)$ 
increases with time during the pulsed non--equilibrium change 
in the inter--site hopping amplitude from its 
ground state value. 

To interpret the 
calculated
time--dependence  
$\Delta S_z(t)$, 
we calculate  
all photoinduced 
populations 
 $\Delta \langle | i m \rangle \langle i m | \rangle$ 
and 
$\Delta \langle | i \alpha M \rangle \langle i \alpha M | \rangle$ 
of the two  sites as function of time. We note in Fig. (\ref{fig2}) that the population of the 
$M$=S+1/2 site 1 ground state 
configuration 
decreases with time due to the laser--driven 
electron hopping to site 2, which creates 
a quasi--hole excitation
$|1 \alpha, S+1/2\rangle \rightarrow 
|1 S \rangle$
on site 1. This excitation 
leaves the core spin unchanged 
with  $m$=S, 
so $\Delta S_z(1) \approx 0$.
The electron 
hops to site 2 while conserving spin, 
by creating a
local excitation 
$| 2, -S \rangle \rightarrow 
|2 \alpha, -(S-1/2)\rangle$.
The population of  these 
$M=-(S-1/2)$ configurations on site 2 
 results in
quantum canting of the local spin 
 $\Delta S_z(2) \ne 0$.
For very short dephasing times, 
the
charge fluctuations
terminate 
after the initial charge transfer.
For longer
$T_2$, 
there are   additional populations of other spin states, which 
which further enhances $\Delta S_z$
In this case, the photo--electron can hop back to site 1 
from site 2 before the laser--induced inter--site 
coherence is destroyed, which 
can lead  to non--equilibrium molecular bonding 
between AFM sites  
that modifies the inter--site magnetic exchange interaction.   
In all cases, 
charge photoexcitation 
induces quantum dynamics   of 
$\Delta S_z(t)$, which increases from zero 
during electron hopping  
(quasi--instantaneous FM correlation). 
 $\Delta S_z(t)$ describes 
non--adiabatic dynamics of the background spins, 
which leads to spin--canting from the quasi--equilibrium 
directions $\theta_i$ that is 
driven by the photoexcitation of quasi--particles dressed by spin fluctuations as discussed next. 

\section{Dependence of Itinerant Quasi--Particle Bands on Spin Fluctuations} 
\label{dispersion} 

In this section we discuss the 
effects of changes in the
lattice potential 
and  spin canting  induced 
by the laser excitation and their possible implications for a 
non--equilibrium phase transition. 
Similar to previous works, 
we  assume that the lattice 
motion  can be described by 
classical coordinates $Q$ \cite{Dagotto,Krish,Ramakrish}.
The 
eigenstates $| i \alpha M \rangle$ of the local Hamiltonian depend on $Q$.  
We model the effects of electron--lattice 
coupling by introducing a
linear Q--dependence  of the 
energy eigenvalues  $E_{i}(\alpha M)$ 
similar to Refs. \cite{Dagotto,tanaka} and neglect any  changes in the 
hopping parameters, which are  
 less known in the
real materials \cite{Loktev}.

Eq.(\ref{HF-coh}) 
suggests a Hubbard--I approximation 
for describing the itinerant 
quasi--particles 
 \cite{Hubbard1,Hubbard2}:
\begin{eqnarray} 
&& \omega_n 
\, \hat{e}_{\alpha \sigma}(i)   
=\varepsilon_{\alpha \sigma}(i)  
\, \hat{e}_{\alpha \sigma}(i )  
\nonumber \\
&&
 -   n_{\alpha \sigma}(i)
\, \sum_{l} 
\sum_{\beta^{\prime}}
V_{
\alpha \beta^\prime}(i-l) \ 
\left[ \cos\left(\frac{\theta_{l} - \theta_{i}}{2} \right) 
 \hat{e}_{ \beta^{\prime} \sigma}(l, t) 
- \sigma 
\sin\left(\frac{\theta_{l} - \theta_{i}}{2} \right) 
\hat{e}_{ \beta^{\prime} - \sigma}(l , t) 
\right].
\label{eom-e} 
\end{eqnarray} 
Introducing  the normal modes 
\begin{equation} 
\hat{e}_{n} =
 \sum_{i \beta \sigma} 
u^{\sigma}_{n}(i \alpha)      
\frac{\hat{e}_{\sigma}(i \beta)}{\sqrt{n_{\beta \sigma}(i)}},
\label{expand} 
\end{equation}
where $n$ labels the different quasi--particle branches (bands), we obtain  
 from Eq. (\ref{eom-e})  the  following eigenvalue equation that gives  
 the itinerant quasi--particle energy bands: 
\begin{eqnarray} 
&& \left( \omega_{n} - \varepsilon_{\beta \sigma}(j) \right) u^{ \sigma}_{n}(j \beta)
=
- 
 \sum_{l \alpha }  V_{\alpha \beta}(l-j)\,  \sqrt{n_{\beta \sigma}(j)} 
\, \sqrt{n_{\alpha \sigma}(l)}
\,
\cos\left(\frac{\theta_{l} - \theta_{j}}{2} \right)
 u^{\sigma}_{n}(l \alpha)
 \nonumber \\
&&
+\sigma  
 \sum_{l \alpha }   V_{\alpha \beta}(l-j) \, 
\sqrt{n_{\beta \sigma}(j)} 
\, \sqrt{n_{\alpha -\sigma}(l)} \, 
\sin\left(\frac{\theta_{l} - \theta_{j}}{2} \right) 
 u^{-\sigma}_{n}(l \alpha).
\label{u-eigen}  
\end{eqnarray}  
The above eigenvalue equation  couples 
 $\sigma$=1 (parallel quasi--particle spin) 
and $\sigma$=-1 (anti--parallel quasi--particle 
spin) excitations (second line). 
By neglecting this coupling,  
the $\sigma$=1 contribution
recovers the previous results for  
bare electrons and parallel 
classical spins 
 \cite{anderson,Krish,Dagotto,Brink}. 
In this case,
the quasi--particle spin 
is locked parallel to the background spins
throughout the motion and  
the hopping 
amplitudes 
$V_{\alpha \beta}(l-j)\, 
\sqrt{n_{\beta \uparrow}(j) n_{\alpha \uparrow}(l)}
\cos\left(\frac{\theta_{l} - \theta_{j}}{2} \right)
$
are maximized  between 
FM sites with $\theta_i$=$\theta_j$. 
On the other hand, for quantum spins,
the finite  
$n_{\alpha \downarrow}(i)$
couples $\sigma$=1 and $\sigma$=-1 excitations and 
allows  composite fermion quasi--particles 
to hop
between AFM sites
$|\theta_l - \theta_j| \sim \pi$.
In this way, 
the quantum spin fluctuations 
couple  the AFM chains and planes 
of Fig. (\ref{fig1})(a) (second term on the rhs of Eq.(\ref{u-eigen}))
and lead to quasi--particle delocalization that strongly affect the energy bands. 
The results of this calculation are shown in Fig. (\ref{fig5}), which compares the effects of quantum spin fluctuations with the energy bands obtained 
in the case of an adiabatic spin background that does not deform during the electronic motion 
(classical spin limit). The  results  of Fig. (\ref{fig5}) were obtained 
with a real space calculation
of a system with periodic boundary conditions, 
which converges for sufficiently large system size  
and reproduces the results obtained 
by assuming a  periodic lattice  \cite{}. 
We consider the CE--AFM unit cell, shown in Fig. \ref{fig1}(a),
with two AFM--coupled zig--zag chains 
that consist of two JT--distorted bridge sites 
and two undistorted corner sites with two 
orbitals \cite{Dagotto}. 
Two AFM--coupled planes as in Fig. \ref{fig1}(a) 
have identical charge, lattice, and orbital configurations. 
While the \lq\lq{valence band}\rq\rq{} that is occupied in the insulating ground state 
is not strongly affected by quantum spin 
flucutations, Fig. (\ref{fig5}) shows that  the \lq\lq{conduction band}\rq\rq{} 
broadens significantly as a result of the added 
electronic delocalization between AFM chains and planes that is suppressed for classical (non--defomrable) spins. 
Eq.(\ref{u-eigen}) thus describes \lq\lq{soft}\rq\rq{} energy bands
of itinerant composite fermion quasi--particles,  
which 
depend non--perturbatively 
on the local spin and charge  populations
of the different lattice sites.
 In addition, the quasi--particle 
energy bands depend on the 
background 
 spins via the classical spin--canting 
 angles $\theta_i$, 
whose effects are shown in Fig. (\ref{fig5}).   
 Finally, the energy bands depend 
 on the classical lattice displacements $Q$ and their dynamics, which determine the
 eigenstates $| i \alpha M \rangle$  on each lattice site and lead to lattice--dependent 
 local excitation energies $\varepsilon_{\alpha \sigma}(i)$, Eq. (\ref{e-energ}). 
 Fig. (\ref{fig5}) shows that the \lq\lq{valence band}\rq\rq{} depends 
 strongly on the JT energy gap, unlike for the conduction band. 
 As a result, the insulating gap increases with JT lattice deformation but decreases due to spin canting, which results in two competing effects 
 that affect differently the ``conduction" and ``valence" bands.
 In the next sections we discuss the possible role of such 
\lq\lq{soft}\rq\rq{} quasi--particle energy bands  in photo--induced phase transitions.

\subsection{Dependence of Quasiparticle Excitations on Lattice and Spin Dynamics}

\begin{figure}
   \begin{center}
   \begin{tabular}{c} 
   \includegraphics[scale=0.55]{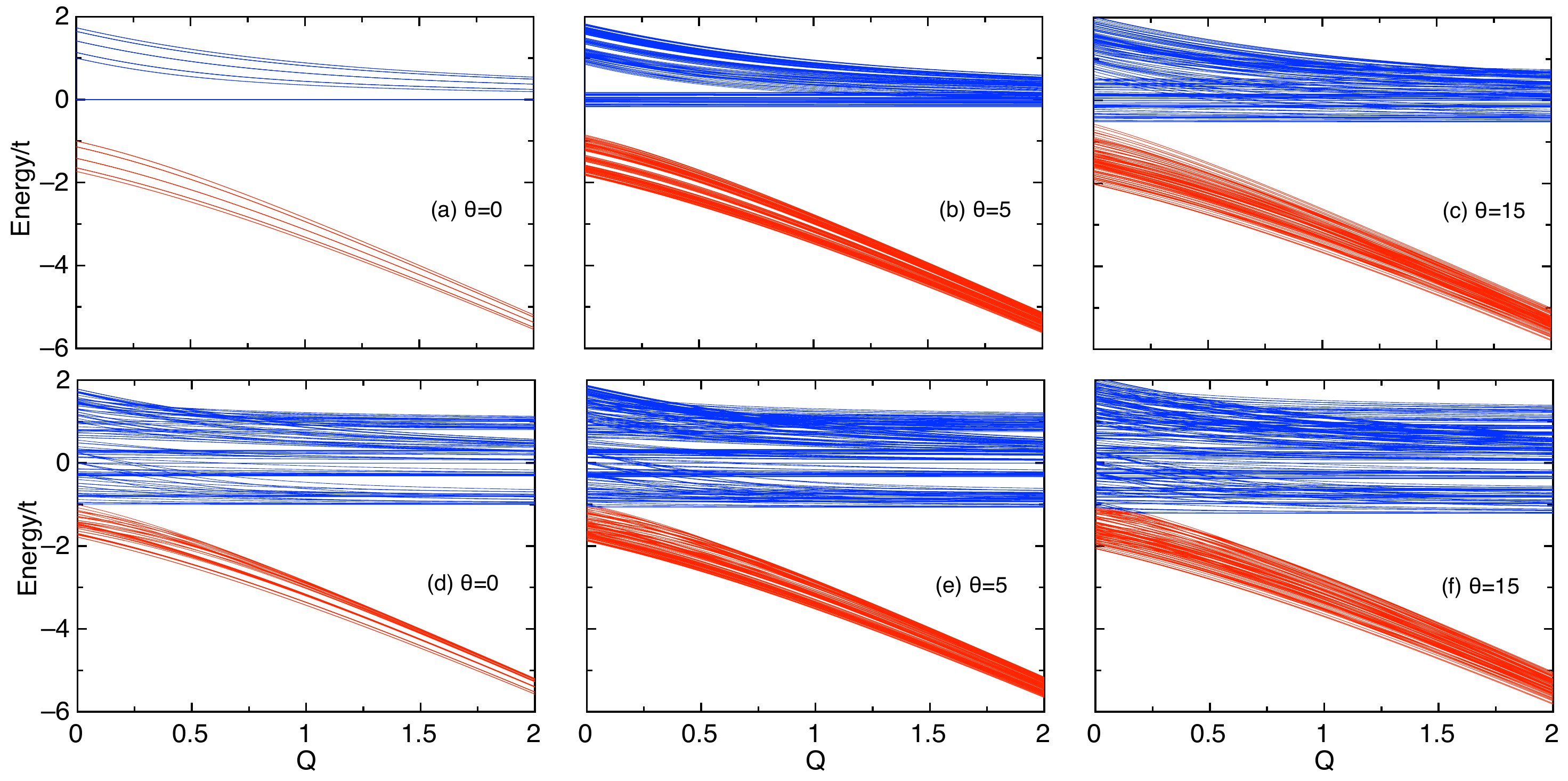}
   \end{tabular}
   \end{center}
   \caption
   { \label{fig5} 
Calculated energy levels  
as function of lattice displacement, 
$E_{JT}$=$\varepsilon_{JT} Q$, 
for  different spin canting angles  $\theta$, which describe FM correlation   
between the FM chains and planes with respect to the collinear AFM state $\theta$=0.  
Upper panel: Bare electrons (adiabatic spins, $S\rightarrow \infty$).  
Lower panel: Composite  fermion quasi--particles (strongly--coupled quantum spins, $S$=3/2). 
$\varepsilon_{JT}=2.5 t_0$. }
   \end{figure}

Fig. (\ref{fig5}) shows the 
dependence of the low energy quasi--particle energy bands  
on the JT energy barrier
$E_{JT}(Q)=\varepsilon_{JT} Q$
between JT--distorted bridge sites and 
undistorted corner sites, which populate 
zig--zag AFM--coupled  chains in neighboring 
AFM planes \cite{}. 
We compare the band Q--dependence 
between  bare electrons, which  move on top of an  adiabatic classical spin background (upper panel), 
and composite fermion quasi--particles (lower panel), whose motion deforms the background spins. 

The main feature demonstrated by Fig. (\ref{fig5}) 
is that the conduction and valence bands 
have different dependence on both  lattice distortions 
and spin fluctuations. 
As seen in Fig. (\ref{fig5}), the energy gap is smaller 
in the case of deformable quantum spins, while it increases with lattice distortion $Q$.  
In the ground state, the 
system is insulating, 
 so $Q > Q_c$ is required in the case of quantum spins,  with $Q_c$ been the critical displacement where the band gap closes. 
For classical spins (adibatic non--deformable spin background), the upper panel of Fig. (\ref{fig5}) reproduces previous results \cite{Krish}. 
For collinear CE--AFM order $\theta$=0,  in this case the 
energy gap does not close even for undistorted lattice $Q$=0 
 due to the  electronic order  
of a single  zig--zag chain \cite{Krish,Dagotto}.   
In the case of classical spins, Fig. \ref{fig5}(a) shows that the lowest conduction band state  is 
a discrete degenerate state with energy 
$\varepsilon=0$.
The latter is 
a linear superposition of the 
electronic configurations in the two different corner sites of the zig--zag chain unit cell and thus does not 
depend on lattice deformation $Q$ \cite{Krish}. 
This is unlike for the valence band, 
which has strong contributions from the 
JT--deformed bridge sites.
With increasing spin canting angle $\theta$ between the AFM
 chains,  electronic hopping between planes 
 breaks the degeneracy of the above 
$\varepsilon$=0 state, which broadens into the lowest conduction band \cite{Krish}. 
For large FM correlation between the chains, $\theta \sim 15^{o}$, the energy gap closes, which results in metallic behavior induced by spin canting.  

To compare with the above classical spin results, 
the lower panel of Fig. (\ref{fig5}) shows that, with $n_{\downarrow} \ne 0$, 
the conduction band  of composite fermion 
quasi--particles is already very broad and metallic in the collinear AFM ground state $\theta$=0.
Such metallic conduction band   
arises from the quantum spin canting 
induced 
by  the  excitation of quasi--electron
population in the conduction band that is mostly 
empty in the insulating AFM equilibrium state. 
Such conduction electrons 
can tunnel between the different 
AFM planes and chains 
due to electron--magnon quantum fluctuations 
that cant the background spins. 
The insulating state requires a finite JT lattice displacement $Q \ne 0$ to obtain a finite 
excitation energy gap. 
In this case,  the valence band  is full in equilibrium while the conduction band is mostly empty. Therefore, quantum spin fluctuations are 
not very pronounced in equilibrium, as can be seen by the small overall difference in the valence band 
between classical and quantum spins in the above figure. 
While  treating the spin background 
as 
adiabatic 
assumes that it is   
slower than the  electronic hopping,
for  composite fermions  quantum spin canting 
occurs during electronic hopping timescales.
This results in instantaneous 
metallic behavior and  FM 
correlation 
during quasi--particle excitation even for large $Q$, which, 
as seen in Fig. \ref{fig5}(d), can  
 quench the energy gap prior to any increase in $\theta$.
With increasing FM correlation between the chains, $\theta>0$, the value of the critical  lattice distortion $Q_c>0$ below which the energy gap closes increases. 
Figs. \ref{fig5}(d), (e), and (f) suggest 
 that an insulator 
to metal transition 
will occur when  $Q \le Q_c$.

 \begin{figure}
   \begin{center}
   \begin{tabular}{c} 
   \includegraphics[scale=0.6]{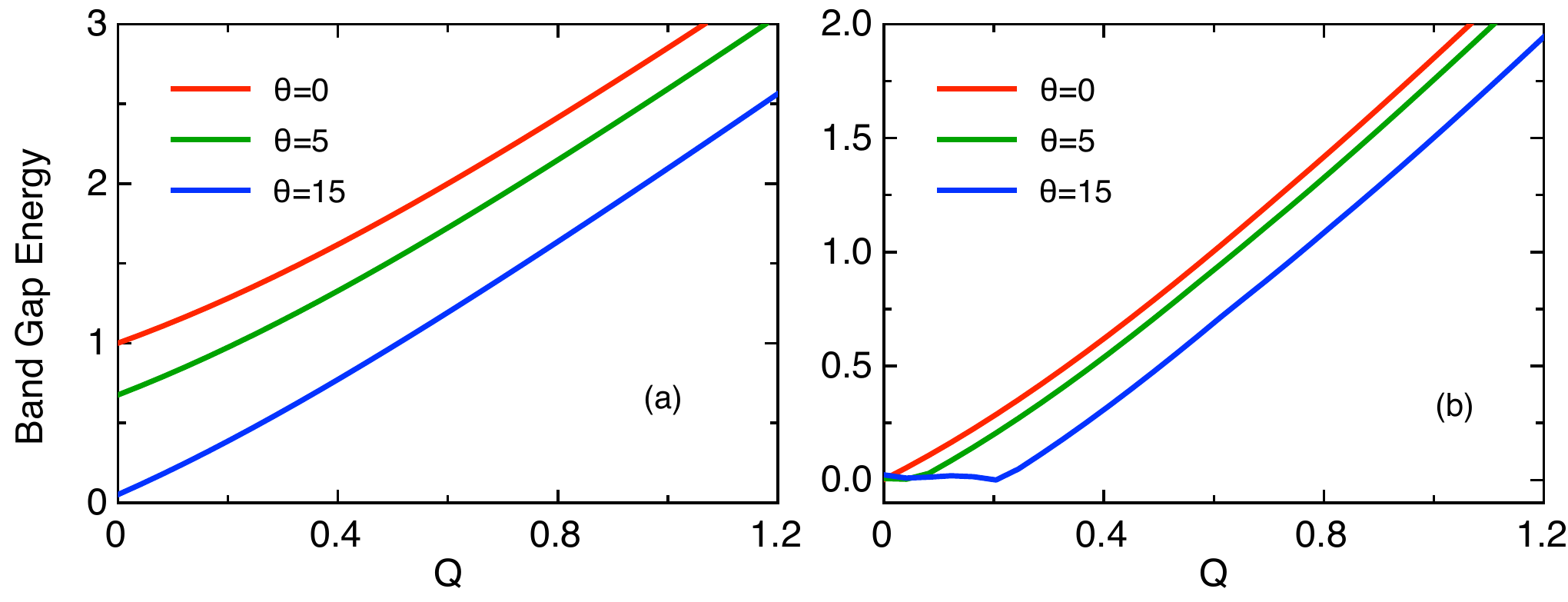}
   \end{tabular}
   \end{center}
   \caption
   { \label{figBG} 
   Effect of  
  spin canting
  on the lattice dependence of the energy 
  gap of Fig. (\ref{fig5}) for increasing $\theta$. (a) bare electrons, 
  (b) composite fermion quasi--particles.}
   \end{figure} 


In the case of composite fermions, 
the charge excitation gap 
depends on the 
spin fluctuations  around the 
average direction  $\theta_i(t)$, 
$\Delta J_z(i) = \Delta S_z(i) 
+ \Delta s_z(i)$, given by Eqs. (\ref{DS}) and (\ref{Ds-itin}).
The latter spin fluctuations can be induced by nonadiabatic photoexcitation of 
local spin populations with 
$M \le S-1/2$ and $m \le S-1$, 
which results 
in time--dependent  changes 
of the composite fermion anti--commutator
Eq.(\ref{factor}) and to quasi--instantaneous non--perturbative 
 changes in the quasi--particle properties and energy dispersion, obtained from Eq. (\ref{factor}):  
\begin{eqnarray} 
 && \Delta n_{\alpha \sigma}(i)=
- \frac{\sigma  \Delta J^\alpha_z(i)
+ \frac{1 - \sigma}{2} \Delta f_i^\alpha
}{2 S + 1}
\label{Dn}
\end{eqnarray}

Fig. (\ref{figBG})  demonstrates an important difference in the Q--dependence of the $e$--$h$ 
quasi--particle energy gap, extracted from Fig. (\ref{fig5}),
between composite fermion and bare electron excitations. 
In the case of bare electrons, Fig. \ref{figBG}(a), 
the energy gap does not close down to $Q=0$
without a large 
 canting angle $\theta$ between the 
 AFM chains.
On the other hand, for composite fermions, Fig. \ref{figBG}(b)  shows that quantum spin canting 
during quasi--particle excitation
softens the energy gap, which now    closes
 below  
$Q$=$Q_c\ge 0$ even for $\theta$=0. 
Fig. (\ref{figBG}) demonstrates the effect of the classical spin canting, which can be induced, e.g., by an external 
magnetic field or by fs laser excitation and results in an increase in the critical lattice displacement $Q_c$ needed to maintain an insulating state.  
The 
critical value $Q_c$  increases with background 
spin canting $\theta$
for both classical and quantum spins. 
However, in the case of bare electrons and assuming  adiabatically--decoupled slower background spins and 
thus more \lq\lq{rigid}\rq\rq quasi--particle bands, 
a  large  spin canting angle $\theta$ is required for $Q_c$ to be 
comparable to that of composite fermions with ``soft'' energy bands. 
As a result, the insulating state is less robust and rigid in the case of 
itinerant electrons moving on top of a deformable spin background (quantum spins). 

\begin{figure}
   \begin{center}
   \begin{tabular}{c} 
   \includegraphics[scale=0.55]{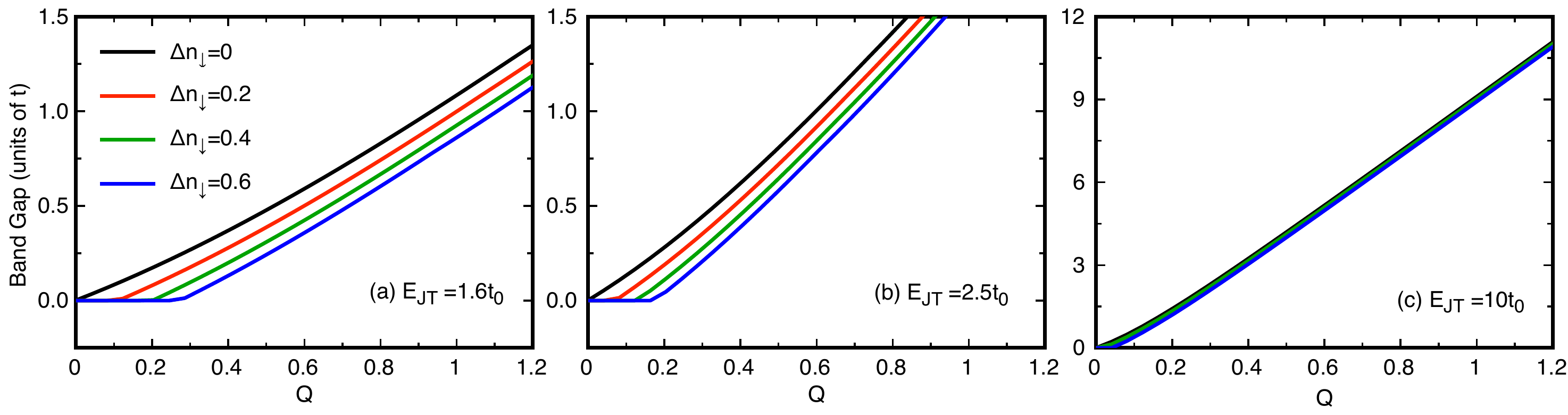}
   \end{tabular}
   \end{center}
   \caption[example] 
   { \label{fig_dn} 
   Effect of  
  FM correlation 
  on the lattice dependence of the energy 
  gap. Composite fermion quasiparticles, increasing $\Delta J_z$. (a) $\varepsilon_{JT}=1.6 t_0$,
  (b) $\varepsilon_{JT}=2.5t_0$, (c) $\varepsilon_{JT}=10t_0$ }
   \end{figure}

As discussed in the previous section, 
laser excitation can result in photoinduced 
FM correlation $\Delta J_z(i) \ne $0.
In Fig. (\ref{fig_dn}) we examine the effect of such spin fluctuations on the  energy  band gap for  various values of the JT splitting between 
JT--distorted and undistorted sites.
The dependence of the composite fermion anti--commutator 
$n_{\alpha \downarrow}(i)$
on the spin fluctuations  $\Delta J_z^\alpha(i)$, 
which can be introduced by the photoexcitation,  
results in a dynamic change of  the 
\lq\lq{soft}\rq\rq{} quasi--particle energy bands
and the insulator bandgap already  
during time evolution of the 
spin populations and local density matrix, 
determined  by 
the equations of motion Eqs.(\ref{dm-XM-1}) and 
(\ref{dm-Xab-1}). 
In all cases, 
photoexcitation of  $\Delta J^{\alpha}_z(i) > 0$ increases $n_{\alpha \downarrow}(i)$, 
which 
instantaneously quenches the energy gap 
and increases 
$Q_c(t)$ as shown in Fig. \ref{fig_dn}(a). Finally, by comparing the results for decreasing JT energy barrier
in Fig. \ref{fig_dn}, we see that 
the effect of spin fluctuations on the itinerant quasi--particle motion diminishes as the JT splitting increases and becomes negligible in the deep insulating limit as shown in Fig. \ref{fig_dn}(c). Most importantly, Fig. \ref{fig_dn}(b) shows 
that in the realistic system with \lq\lq{soft}\rq\rq{} quasi--particle bands and JT energy barrier comparable to the electronic hopping energy, 
quantum spin fluctuations 
 significantly affect the metal--insulator 
 transition, which now requires larger 
 lattice displacements following FM spin photoexcitation.

We conclude that, independent
of the details of fs spin photogeneration,  both 
adiabatic $\theta(t) > 0$ and nonadiabatic  $\Delta J_z(t)>0$
 FM correlation induced by the 
 photoexcited quasi--particles enhances the critical 
 JT displacement $Q_c(t)$
required  for insulator to   
 metal transition
when the lattice displacement $Q(t) \le Q_c(t)$. 
This change in $Q_c(t)$ may   already 
occur during photoexcitation of 
sufficient population of composite fermion quasi--particles, 
which leads to instantaneous FM correlation 
due to  quantum spin fluctuations that 
quenches the insulator electronic energy gap.  
The  detailed  time--dependence of the 
photoinduced spin canting  in different materials, 
which determines the critical photocarrier density such that $Q(t)<Q_c(t)$ as required for 
 a phase transition, 
depends on the details of the material, which are not well known. 
In all cases, 
Figs (\ref{figBG}) and (\ref{fig_dn}) imply a nonlinear dependence
of the electronic properties 
 on  the pump fluence, as the latter  
controls the non--thermal populations 
of the composite fermion 
 excitations that \lq\lq{suddenly}\rq\rq{} change  the \lq\lq{soft}\rq\rq{} energy bands and  $Q_c(t)$, 
while  also inducing  lattice displacements $Q(t)<Q$
 as discussed next. We expect that, with multiple quasi--particle
excitations,  
such nonlinear dependence will be even stronger than the single quasi--particle 
results presented here.
 We also note that classical spin equilibrium calculations \cite{Krish} predict a very high critical magnetic  
field for inducing a CMR phase transition, due to the large charge excitation energy gap
for adiabatically--decoupled spins. 
Here we argue that composite fermion excitations characterized by  \lq\lq{soft}\rq\rq{} energy bands that can be manipulated 
optically can make an insulator to metal and AFM to FM 
transition possible for low magnetic fields and pump fluences. 
 
\subsection{Photoinduced softening of the lattice displacement}

As discussed in the previous section, 
photoexcitation can induce an insulator--to--metal transition by changing the lattice 
displacement ib equilibrium, $Q > Q_c$, to 
$Q(t) \le Q_c(t)$. This can 
be achieved  
  in two different ways: 
(i) increase critical displacement $Q_c(t)>Q_c$, (ii) decrease in lattice displacement 
$Q(t)<Q$.
In this section we explore the second possibility during 
nonthermal fs timescales. 
The effective 
potential  
that governs  the 
lattice motion leading to  $Q(t)$ 
includes both 
 the classical contribution 
 $U_{L}(Q)$, which can be obtained phenomenologically 
 based on the symmetry \cite{landscape,Beaud}, and the contribution 
of the local electron--lattice coupling. 
The latter is  important for the  laser--induced phase transition 
 proposed here and  is 
described by the Q--dependence of the Hamiltonian 
Eq.(\ref{H(t)}). 
The lattice equations of motion
can be derived  as in Ref. \cite{TD-TB,Lingos_PRB_2017}.
For this we introduce an 
orthogononal  basis of many--electron ground state $|G\rangle$ 
excited states 
$| E \rangle$ of $H(Q)$ and expand the time--dependent many--body 
state  
 evolving from the equilibrium state $|G\rangle$
following photoexcitation.  
The time--dependence of the  lattice coordinates 
is described by 
$M_l \frac{d^2 Q_l}{d t^2} = 
F_l(Q)$, where the forces are 
determined by  the  
 time--dependent populations $f_E(t)(t)$ of the above many--body eigenstates 
\begin{eqnarray}  
F_l (Q,t) \approx - \frac{\partial}{\partial Q_l} \
 \left[ U_L(Q) + \langle G| H(Q)| G \rangle  \right]  - \sum_{E \ne G}
f_E(t) \
\frac{\partial \varepsilon_{E}(Q)}{
\partial Q_l}, 
\label{force}
\end{eqnarray} 
where 
\begin{equation} 
\varepsilon_E(Q)= \langle E| H(Q) |E \rangle  
- \langle G| H(Q) |G \rangle
\label{gap}
\end{equation} 
are the $e$--$h$ quasi--particle excitation energies. The latter are described by  
Fig. (\ref{fig5}) and depend on the lattice displacement $Q(t)$.   
The first term on the rhs of Eq.(\ref{force}) 
gives the adiabatic potential, which determines the 
lattice motion 
in the case of adiabatic time evolution 
of the insulating ground state $|G \rangle$
of $H(Q)$
 without quasi--particle excitation \cite{subedi}. 
The second term on the rhs of Eq.(\ref{force})
describes a quasi--instantaneous  change in the lattice potential and forces 
 when the population of excited many--body states $| E \rangle$
becomes significant.  
Such non--equilibrium potential change initiates lattice motion 
following quasi--particle excitation 
and changes with time 
 as determined by the evolution of the 
non--equilibrium populations 
 $f_{E}(t)$
 and by the dependence of the
 excitation energies Eq.(\ref{gap}) on $Q(t)$. 
This is analogous to  previous results 
in VO$_{2}$ \cite{Veenendaal-VO2,He}
and semiconductors \cite{zeiger}.   
A phase transition is triggered if $Q(t) \le Q_c(t)$
during the lattice motion $Q(t)$, which can involve 
both coherent phonon oscillations and/or anharmonic lattice motion. 
Fig. (\ref{fig5}) indicates a nonlinear Q--dependence of both ground state energy and $e$--$h$ 
quasi--particle excitation energies, so 
Eq.(\ref{force}) 
implies that the effective spring constants which 
determine, e.g., the coherent phonon oscillation frequencies,
will change from their quasi--equilibrium values following photoexcitation of excited populations $f_{E}(t)$.
On the other hand, 
 for $E_{JT}(Q) \gg t$, 
 the energy band $Q$--dependence is approximately linear, 
 which implies much smaller changes in the sping constants.  The fs--resolved XRD experimental results of Ref. \cite{Beaud} 
show that the photocarrier density transiently modifies the lattice spring constants in the manganites.

The laser--induced changes in the lattice 
potential and forces
with quasi--particle excitation will initiate a lattice motion  that depends on  both $f_E(t)$ and  
 $\frac{\partial \varepsilon_E(Q)}{\partial Q}$.
New metastable quasi--equilibrium lattice configurations 
$ Q_{eq}(t)$ can be  obtained
from  Eq.(\ref{force}) 
by setting  $F_l(Q)=0$.
Such configurations depend on the elastic potential $U_L(Q)$, 
determined by multiple lattice modes  and lattice anharmonicities \cite{landscape,subedi}.  
For our purposes here, we assume for simplicity that 
$U_L(Q)=\frac{1}{2} k Q^2$. In this case, 
Eq.(\ref{force}) gives quasi--equilibrium lattice 
configurations that depend on the photocarrier density: 
\begin{equation}  
 Q_{eq}(t)=
- \frac{1}{k} 
\frac{\partial}{\partial Q} \langle G| H(Q)| G \rangle   - \frac{1}{k} \sum_{E \ne G}
f_E(t) \
\frac{\partial \varepsilon_E(Q)}{
\partial Q}.
\label{Q-eq}
\end{equation}
The first term determines the equilibrium 
lattice distortions, which 
are modified following photoexcitation of the continuum of 
 many--body states $|E \rangle$. 
 From  
Fig. (\ref{fig5}) we see that  
$\frac{\partial \varepsilon_E(Q)}{
\partial Q} > 0$  is dominated by the hole contribution to the excitation energy. 
The photoexcited quasi--particle populations then decrease the quasi--equilibrium lattice displacements 
to 
$Q_{eq}(t)$ during  timescales where $f_E(t) \ne 0$.  The latter lattice displacement  changes with time as determined by the non--thermal electronic populations, which can lead to time--dependent 
changes in the hoppping amplitudes $\Delta V(t)$
as discussed above.  
The  lattice displacements 
$Q_{eq}(t)$ 
are expected to be small, 
$Q_{eq}(t) \ll Q$, as compared to equilibrium 
 in the case of laser--induced population 
inversion between the two different 
quasi--particle bands of Fig. (\ref{fig5}). 
This is the case as 
the conduction and valence  band eigenstates have different 
admixture of corner and bridge site 
configurations, which leads to their 
different Q--dependence. It may lead to an irreversible transition when 
$Q_{eq}(t) \le Q_c(t)$.

The above  picture of a 
photoinduced insulator to metal transition 
above a critical photocarrier density 
such that $Q(t) \le Q_c(t)$
may be validated by experimental observations of nonlinear and threshold
dependences of the ultrafast 
spectroscopy signals with increasing pump fluence
and with a better temporal resolution that can distinguish between instantaneous 
and time--delayed processes. 
In the non--thermal temporal regime 
of interest here,   a laser--induced population inversion 
between the  polaronic--like 
majority carriers and the  metallic--like 
minority carriers drives a  nonlinear inter--dependence
of spin, charge, and lattice dynamics. 
In this way, fs laser excitation 
can break the balance between electronic/magnetic 
and lattice degrees of freedom based on their different dynamics discussed above.
To test this  experimentally, 
one must be 
able to non--thermally control the quasi--particle populations while simultaneously monitoring the resulting spin, charge, and lattice  
time evolution  on a fs timescale. 
This may be possible by using X--ray pulses \cite{Beaud,Ehrke} as their time resolution improves. 

\section{Conclusions} 
\label{discussion}

To conclude, in this paper we described a possible 
mechanism for  photoinduced insulator to metal 
and AFM to FM simultaneous transitions triggered by 
non--thermal  quasi--particle populations
of a metallic conduction band. 
This mechanism involves non--adiabatic and nonlinear spin--charge--lattice 
coupling in the case of an AFM  ground state consisting of FM chains and planes
with JT distortions that stabilize the insulator energy gap.  
We propose that this mechanism may be relevant 
to explain the 
  nonlinear pump fluence threshold dependencies of both 
magneto--optical (MOKE and MCD)
and $\Delta R/R$ femtosecond signals 
 measured in the PCMO manganite\cite{Lingos_PRB_2017}. 
It may also be relevant   
to several other ultrafast spectroscopy experimental observations 
of nonlinear behavior 
during  the non--thermal temporal regime following fs laser excitation 
of the AFM 
state of different insulating manganites 
\cite{PolliNat,RiniNat,Beaud,Forst,Miyasaka,Matsubara,Okimoto,Wall,Ehrke,Matsuzaki,singla}. 
In particular, we predict that electron--spin correlation 
leads to a broad conduction metallic band 
and quenches the electronic component of the 
insulator energy gap below a critical 
value of the JT lattice displacement that depends on the photoexcitation. 
Such laser--induced effects are pronounced in the 
case of composite fermion quasi--particles 
with ``soft'' energy bands, which mostly populate the lower magnetic Hubbard band due to 
the large Hund's rule interaction and excite spin flucuations that lower their energy during electronic hopping
timescales.  
Photoinduced FM correlation and 
spin canting following composite fermion 
quasi--particle photoexcitation instantaneously increases the critical lattice displacement $Q_c(t)$ 
below which the energy gap closes and  leads to 
a metal--insulator phase transition that is also facilitated by 
the non--instantaneous relaxation of the JT distortions.  
In particular, above a critical photocarrier density, 
$Q_c(t)$ can become comparable 
to the equilibrium lattice distortion, while the latter decreases following lattice motion. 
Both effects act cooperatively to induce 
non--equilibrium insulator to metal and AFM to FM 
simultaneous transitions. 
We showed that FM correlation can develop instantaneously, so it can induce an instantaneous 
insulator to metal transition if $Q_c(t) \ge Q(t)$ 
during the laser pulse. 
Otherwise, lattice relaxation or oscillation 
is also required so that 
$Q(t)$ decreases. 
The excitation of multiple quasi--particles 
should increase the above nonlinearities 
by enhancing the deformation of the AFM background.  
After such excited quasi--electrons
 have relaxed on a fs timescale ($\tau^{\mathrm{fs}}$), 
electron--lattice and spin--lattice 
relaxation   determines 
the subsequent ps 
dynamics 
($\tau^{\mathrm{ps}}$). 

The above theoretical framework
may be relevant for explaining several experimental  observations when worked cooperatively with lattice deformation and free energy quasi--equilibrium effects.
The observation of 
a time--dependent spring constant \cite{Beaud} 
and nonlinear dependence of the coherent 
phonon amplitudes on the pump intensity 
during non--thermal fs timescales 
are consistent with \lq\lq{soft}\rq\rq{} quasi--particle 
energy bands such as the ones proposed here.  
Such bands of composite fermion quasi--particles make it easier to 
obtain a quasi--instantaneous insulator to metal 
and AFM to FM transitions as compared to bare electrons adiabatically decoupled from the spin background 
and nonlinear dependence of the coherent 
phonon amplitudes on the pump intensity 
during non--thermal fs timescales.
In the latter case, 
the mechanism must rely on a more elaborate 
non--instantaneous lattice 
motion \cite{veenendaal} 
in order to close the energy gap. 
In addition, a complex energy landscape, possibly  
with multiple local minima 
due to the elastic lattice potential $U_L(Q)$ 
\cite{landscape,Beaud}, 
should  facilitate the phase transition mechanism 
proposed here by increasing $Q_c$ and creating metastable states. 
The insights  
from  our theory, accompanied with relevant experiment  results\cite{Lingos_PRB_2017}, 
may also prove useful for 
revealing the crucial many--body processes in other intertwined electronic phases, as the proximity of magnetic states appears 
ubiquitous with unconventional superconducting and exotic electronic 
phases in strongly correlated electronic materials \cite{Aaron}.
In the long run,  new insights can be gained  by applying complementary ultrafast spectroscopy techniques, especially in the terahertz \cite{luo2015} and infrared spectral regions \cite{mid-ir}, and by combining spin and charge quantum fluctuations with 
quasi--equilibrium 
free energy and self--energy  effects.

\end{document}